\documentclass[reqno]{amsart}
\usepackage{graphicx,color}
\usepackage{amssymb,amsthm,amsmath,amsfonts}

\newcommand{\pd}[2]{\frac{\partial #1}{\partial #2}}

\newcommand{\fe}{E} \newcommand{\fv}{V} \newcommand{\fs}{S}
\newcommand{\re}[1]{(\ref{#1})}
\def\rdot[#1]{\overset{\underset{\negyzet}{}}{#1}} 
\def\negyzet{\hbox{\tiny$\zS\blacksquare$}}

\newcount\n  \newdimen\w

\def\Repeat#1#2{\n=#1\relax\loop\ifnum       
  \n>0\relax #2\advance\n by-1\repeat}

\long\def\OMIT#1{\relax }  

\def\sect#1#2{{\section{#2}\label{#1}}}

\def\ssect#1#2{\subsection{#2}\label{#1}}

\def\re#1{(\ref{#1})}   
  
\def\eqn#1#2{ \begin{align} \label{#1}         #2 \end{align}}

\def\nnl#1{ \tag*{} \\ \label{#1}        }  
  
   \let\zS\scriptscriptstyle

\def\delim#1#2#3{\csname\ifcase#1 relax\or   
   big\or Big\or bigg\or Bigg\fi\endcsname   
  {\ifcase#2\or\Delim#3\or\deliM#3\fi}}      
\def\Delim#1{\ifcase#1\relax\or(\or[\or\{\or<\or\langle\or|\or\|\or---{ }\fi}
\def\deliM#1{\ifcase#1\relax\or)\or]\or\}\or>\or\rangle\or|\or\|\or{ }---\fi}

\let\f\frac                     
        
\def\largerfrac#1#2#3{      
  \whichtypesize\n=\currenttypesize\advance\n by #1 \mathchoice
  {\setbox0\hbox{$\displaystyle-$} \w=.5\ht0\advance\w by-.5\dp0\setbox0
    \hbox{\typesize\n $\displaystyle-$} \advance\w by -.5\ht0\advance\w
    by .5\dp0\raise\w \hbox{\typesize\n$\displaystyle{\frac{#2}{#3}}$}}
  {\setbox0\hbox{$-$} \w=.5\ht0 \advance\w by -.5\dp0 \setbox0\hbox
    {\typesize\n $-$} \advance\w by-.5\ht0\advance\w by
    .5\dp0\raise\w\hbox{\typesize\n$\frac{#2}{#3}$}}
  {\setbox0\hbox{$\scriptstyle-$} \w=.5\ht0 \advance\w by-.5\dp0\setbox0
    \hbox{\typesize\n $\scriptstyle-$} \advance\w by -.5\ht0 \advance\w
    by .5\dp0 \raise\w\hbox{\typesize\n$\scriptstyle{\frac{#2}{#3}}$}}
  {\setbox0\hbox{$\scriptscriptstyle-$} \w=.5\ht0
    \advance\w by -.5\dp0 \setbox0\hbox{\typesize\n
    $\scriptscriptstyle-$} \advance\w by -.5\ht0 \advance\w by .5\dp0
    \raise\w\hbox{\typesize\n$\scriptscriptstyle{\frac{#2}{#3}}$}}  }

\def\d{{\rm d}}     

\begin{document}
\title{Toward a Universal Theory of Stable Evolution}
\author{P. V\'an$^{1,2}$ }
\address{$^1$Department of Theoretical Physics, Wigner Research Centre for Physics, H-1525 Budapest, Konkoly Thege Miklós u. 29-33., Hungary; //
and  $^2$Department of Energy Engineering, Faculty of Mechanical Engineering, Budapest University of Technology and Economics, H-1111 Budapest, Műegyetem rkp. 3., Hungary\\
ORCID: 0000-0002-9396-4073}
\date{\today}

\begin{abstract}
The backbone of nonequilibrium thermodynamics is the stability structure, where entropy is related to a Lyapunov function of thermodynamic equilibrium. Stability is the background of natural selection: unstable systems are temporary, and stable ones survive. The physical concepts from the stability structure and the related formalism of constrained entropy inequality are universal by construction. Therefore, the mathematical tools and the physical concepts of thermodynamics help formulate dynamical theories of any systems in social and natural sciences. 
\end{abstract}
\maketitle

\section{Introduction}

Universality is a remarkable property of physical theories, quantities and parameters. Temperature, second-order phase transitions, gravity, or the gas constant are universal in the sense that they do not depend on material properties. One of the famous demonstration of the universality of free fall is the Eötvös-Pekár-Fekete experiment. There the measured ratio of inertial and gravitational masses was the same for several different materials \cite{EPF22a}. Since then, many experiments have verified and improved that for different length scales and materials \cite{Wil14a}. For second-order phase transitions, the approach is similar: for different materials and material structures the critical exponents are the same, \cite{Gug45a,Sta71b}. There is also an empirical part of the universality of various physical constants, like the universal gas constant.  

In the experiments, one recognises the independence from material. In the related theories, the empirical facts must be explained. Therefore universality becomes a construction principle, like in the case of general relativity, or can remain unexplained, but is a source of insight and methodological development, like in the case of critical exponents. The gas constant is a built-in property of the related theories, and its universality has somehow become natural. 

The universality of temperature, i.e. the absolute temperature scale, is peculiar. It is the only case where we trace back the independence from matter to more fundamental aspects. Those are the reversibility and composability of processes and also the Second Law, a particular property of thermodynamic engines (see, e.g. \cite{Kar07b1}). "Engines" here are abstract representations of materials. Then the absolute temperature and, consequently, the existence of entropy follow. The argumentation can be and should be extended at least to some other thermodynamic quantities, like the pressure \cite{Gya62a}. The key concept is the strange, fictitious "process" of equilibrium thermodynamics. In equilibrium thermodynamics, processes are time-independent; the change of thermodynamic state refers to so-called quasi-static processes that run through equilibrium states. This is why the scope of thermodynamics is generally thought to be limited. It is a strange contradiction: because of universality, substances have temperature, independently from their material structure, let it be electromagnetic field, fluid or an ensemble of particles, and, at the same time, the starting point of the deductive justification has limited validity. This dilemma about the validity range of thermodynamics is remarkable and thought-provoking (compare  \cite{Atk07b,Ben17b}). 

Another important manifestation of universality is related to matter-related evolution equations, the ordinary or partial differential equations describing the processes. The Fourier equation of heat conduction and the Fick equation of diffusion are prominent examples: these are the same differential equations that are valid for very different materials and mechanisms. The equation works well for heat conduction and diffusion, in fluids and solids as well. The material's structure is only reflected in the parameters of the differential equation; the heat conduction coefficient, diffusion coefficient, and density are different, but the equation is the same. This observation motivated, for example, the research direction called synergetics \cite{Hak83b}. That kind of universality of the evolution equations will be called {\em process universality}. It is mainly related to nonequilibrium thermodynamics: in the particular case of synergetics, it has been linked to order parameters of phase transitions, the fields describing non-equilibrium structures. Nevertheless, that kind of explanation remains at the empirical level: no principle in synergetics could explain the universality of the evolution equations in detail. 

The appearance of thermodynamic concepts and principles in biological and social processes can be another example of empirical universality. One can see - or rather feel -- the applicability, but without a physical or any insight that could explain it. In the case of biological evolution and economics, thermodynamics not only emerges as a helpful analogy but also leads to a useful conceptual framework\footnote{There are too many and too different suggested connections of biological systems and thermodynamics. Classical irreversible thermodynamics focused on the concept of dissipative structures \cite{GlaPri71b}. Some other exciting suggestions are \cite{Gla78a,Bej17a,Ske17a,Swe23m}. It is also remarkable that the mechanism of natural selection is based on a stability argument \cite{Den96b}. In economics, there were also high hopes in classical and irreversible thermodynamics, \cite{Geo71b,Sam89a}, and these approaches were further refined also for its process models \cite{AyrMar05b,Kum11b}. Recent approaches focus on statistical aspects, see, e.g. \cite{SchaYan20a}.}. One may ask what is the reason for the appearance of thermodynamics in this many areas of sciences and humanities and what is its limited success?

This paper argues that thermodynamic state changes of homogeneous bodies can be understood as real processes and that conditions for asymptotic stability of thermodynamic equilibrium are based on concepts of thermostatics. In this framework, i.e., for non-equilibrium thermodynamics of homogeneous bodies, which can be called ordinary thermodynamics, the total entropy is the Lyapunov function that ensures the stability of equilibrium. Thus, equilibrium thermodynamics is a dynamical theory in disguise, where the stability properties of the dynamics interpret the second law and entropy. Hence, a conceptual background emerges for the above-mentioned process universality. The stability aspect is why thermodynamics appears in so many areas, and it is the key to extend  its applicability. 

This paper explains the stability structure of equilibrium thermodynamics in the case of the simplest thermodynamic system. Then the universality of temperature is analysed in the light of the stability background, and, in the end, the scope of the stability based approach to social and scientific dynamics is shortly discussed.

\section{Second Law and Lyapunov stability}

Classical equilibrium thermodynamics (hereafter called \textit{thermostatics}) developed during the 19th century. By this time, mechanics was the basis of natural philosophy and the model for all other new physical disciplines. However, thermostatics, the newest of classical branches of physics\footnote{Only the second and the first laws were recognised then, the third law was formulated in 1912, zeroth law in 1939 \cite{FowGug39b}.} is not a dynamical theory, and it is not based on an evolution equation. Non-equilibrium thermodynamics appeared as a classical field theory, its evolution equations describe the evolution of continua in space and time. A dynamic theory of homogeneous bodies would be a natural reduction, which was attempted in the monograph of de Groot and Mazur, \cite{GroMaz62b}, among others. However, the homogenisation of the continuum equations of non-equilibrium thermodynamics did not yield a theory compatible with thermostatics. Subsequently, the research of Truesdell and Bharatha showed that the other, bottom up construction is not straightforward either, it is not obvious to complement the first law with an additional differential equation and to keep the structure of thermostatics at the same time, \cite{TruBha77b}.    

Further difficulties in the non-equilibrium thermodynamics of homogeneous bodies are illustrated by the later developments, by the \emph{finite time}, and \emph{endoreversible} thermodynamics \cite{CurAhl75a,And83b,Bej96a}, and also by thermodynamics of discrete systems, \cite{Musch22a}. There time-dependent, real processes are treated without evolution equations and a dynamic interpretation of the Second Law, \cite{Lav10b,Mor98a,Gyf99a}. 

The problem is the connection between the dynamic and static aspects of entropy. The dynamic part of the Second Law is related to insulated systems, where the entropy is expected to grow, but nothing happens in insulated homogeneous bodies; time-dependent processes are only in open thermodynamic systems. The solution of the paradox and the consequent complete dynamic formulation of thermodynamics of homogeneous bodies was given by Matolcsi, \cite{Mat92a1}, and is called {\em ordinary thermodynamics}, \cite{Mat05b}. 

In the following, a brief presentation of the core of ordinary thermodynamics shows that thermodynamic stability, zeroth law and increasing entropy are the conditions for asymptotic stability of thermodynamic equilibrium, and, most remarkably, the total entropy of the thermodynamic system is a Lyapunov function of the thermodynamic equilibrium. This paper's purpose is not to critique thermodynamic theory or to resolve its paradoxes in detail. These are dealt with in many respects in the books mentioned above \cite{TruBha77b,Mat05b,Mat12b}. 

\ssect{1ht_one}{Thermostatics and thermodynamics}

The key concept of thermodynamics is \textit{equilibrium}. However, since there are no real processes in classical theory of homogeneous bodies, and the possible process terms used -- quasistatic, reversible, or irreversible -- are vaguely defined, the concept of thermodynamic equilibrium is complex\footnote{Actually, it is a confusion.}. Nevertheless, thermodynamic quantities are taken to be meaningful only in equilibrium in all introductory thermodynamics books \cite{Cal85b,Rei98b}. Equilibrium is implicitly defined; its meaning can be, among other things, time independence, homogeneity and lacking dissipation. The zeroth law of thermodynamics is intended to clarify the conditions.

The concept of thermodynamic equilibrium is formulated here self-consistently, without any microscopic, statistical background. In this conception, according to the zeroth theorem, equilibrium means the \textit{reducibility} and \textit{separability} of physical systems. Reducibility, since we compress the combined action of many atomic, molecular, and mesoscopic physical quantities into some thermodynamic variables. Separability means that individual physical quantities, the state variables, characterise thermodynamical bodies, even if they are interacting. 

Time independence is not an essential element of the concept of thermodynamics. The relationship between thermodynamics and thermostatics can be clarified understanding the dynamic role of thermodynamic potentials, especially entropy. 
	
To describe the time variation of thermodynamic quantities of homogeneous bodies, we can use ordinary differential equations, the \textit{evolution equations} of the given variables. Since we will be dealing with mechanical systems -- fluids -- these evolution equations are sometimes related to motion, therefore, the evolution equations of thermodynamic theory must incorporate the corresponding form of Newton's equation, both dissipative and nondissipative. In other words, a sufficiently complete theory of non-equilibrium thermodynamics incorporates inertial effects.

\sect{hszt1}{Termostatics}
	
Three basic hypotheses are postulated. They are not mathematical axioms because a complete mathematical precision would obscure the physical background, but some precision is unavoidable for clarity\footnote{For example, I will assume all functions to be differentiable, invertible, etc., although we know well that the discussion of potentials or phase boundaries, requires fixing the differentiability and also the domains of the functions.}. First of all, the existence of a thermodynamic body as a separate entity of reality is a fundamental assumption. 
	
{\it {\bf A1:} There are independent thermodynamic bodies. Extensive state variables and intensive state functions characterise them}.
	
This is the static part of the zeroth law.
	
A thermodynamic body is characterized by the state space consisting of $N$ {\it extensive thermodynamic quantities}, $(X^1,X^2,...,X^N)$ which form a vector space $\mathbb{X}$ and $N$ {\it intensive thermodynamic quantities}, a state function $(Y_1,Y_2,...,Y_N):\mathbb{X}\rightarrow\mathbb{X}^*$, where $\mathbb{X}^*$ is the dual space of $\mathbb{X}$. The components of the intensive state function are experimentally given. The most important assumption here is separability, i.e. thermodynamic bodies can be characterised by their own independent state functions even when they are apparently under the influence of other thermodynamic bodies or the environment due to physical interactions or conservation laws\footnote{The additivity or a more general concept of composability is not trivial either, also with considering the simplest microscopic constitution \cite{BirVan11a}. Additivity is not to be confused with extensivity.}.
	
A defining property of extensive state variables is that they are proportional to the extension of the thermodynamic body. However, the conditions of extensivity are not treated here; they are simply the canonical set of state variables.
	
{\it {\bf A2:} The intensive state function has a potential, the potential is the entropy}.
	
That is, for a thermodynamic body, there exists an entropy function $S:\mathbb{X} \rightarrow \mathbb{R}$, whose derivative is $(Y_1,Y_2,...,Y_N)$. Traditionally this is expressed by the Gibbs relation
\eqn{Gibbs_rel}{
	\d S = Y_A \d X^A = Y_1\d X^1 + Y_2 \d X^2 +...+ Y_N \d X^N.
}
which means more precisely that 
\eqn{int_pardef}{
	\f{\partial S(X^1,X^2,...,X^N) }{\partial X^A}= Y_A(X^1,X^2,...,X^N)
}
Here upper index indicates vector, the lower index covector, and the double index denotes duality mapping. This requirement imposes conditions on the experimentally defined functions: the derivative of the intensive state function is symmetric\footnote{The potential conditions, the so called Maxwell relations, are rarely measured directly. We accept the potential property and design the experiments accordingly.}.
	
{\it {\bf A3:} The entropy is concave}.
	
This requirement is called {\em thermodynamic stability}. Then it follows that the second derivative of the entropy, a second order tensor, is negative definite\footnote{Actually it is only semidefinite if Euler homogeneity is required, its kernel is spanned by all $(X^1,X^2,...,X^N)$, where thermodynamic stability holds, multiplied by a nonnegative number, \cite{Mat05b}.}. 
	
The above conditions are gradually weaker from a physical point of view. Phase transitions and boundaries violate thermodynamic stability, property {\it A3} is local. The existence of macroscopic entropy, defined in {\it A2}, is rarely questioned, mainly in the case, when it is calculated assuming ideal theoretical microstructures. {\it A1} is the most profound assumption, the separability and the possibility of reduced complexity are rudimentary. Theories that would violate {\it A1} are not only not thermodynamic, they are not physical: this kind of reductionism is fundamental in existing theories of physics.
	
In what follows, it will be shown in detail, using the simplest examples, that the above fundamental properties have a common role, they are conditions ensuring the asymptotic stability of equilibrium of thermodynamic systems within the framework of a dynamical theory. This, in turn, highlights how and in what sense thermodynamics is fundamental and why it is part of all physical disciplines: the existence of asymptotically stable states of matter is essential for experimental reproducibility and, thus, for the existence of objective natural science. 

\sect{hszt2}{Termodynamics}
Regarding dynamics, the fundamental assumption is that there is a dynamical law, a differential equation that defines and determines processes. We do not specify that in the form of a formal postulate, a  representative example will be given in the next section. This subsection aims to clearly separate the two independent parts of the Second Law and emphasize that there is an evolution equation in the background.  

{\it {\bf A4:} A system is a collection of interacting bodies, the processes of the system are the solutions of an evolution equation.}

The evolution equation is an ordinary differential equation, but it is not arbitrary:

{\it {\bf A5:} A requirement is imposed on the dynamical equation which ensures that the entropy is an increasing function along the processes in an insulated system. }

Precisely this means that a process of a body is a function defined in time $t\mapsto (X^1(t),X^2(t),...,X^N(t))$, and the entropy change along the process is $t\mapsto S(X^1(t),X^2(t),...,X^N(t))$. Insulated is a physical condition, expressed by the conservation of extensive quantities in the particular cases. The emphasis is on the adjective "along the processes", which are determined by the evolution equation. A demonstrative example and a short analysis of some traps using $A1-A5$ in modeling are given in the following section.

\sect{hgf1}{Ordinary thermodynamics - homogeneous fluids}

The processes of thermodynamic bodies are functions of time in ordinary thermodynamics. In this respect, it is analogous to the mechanics of mass points.  However, in thermodynamic state space is not related to spacetime, space does not play a role at all. Bodies of ordinary thermodynamics are models of matter with homogeneous distribution in space. 

\ssect{hgf1_sztat}{Thermostatics of fluids}

In gases and liquids, the classical extensive state variables are the internal energy \textit{E}, the volume \textit{V} and the particle number \textit{N} (or mass \textit{M}). Entropy is a function of these: \(S(E,V,N)\). 

According to $A2$, entropy is a thermodynamic potential, and its partial derivatives are the entropic intensive state functions. Therefore, if equations of state, the intensive state function, is given, entropy is defined as a potential in terms of its partial derivatives : 
\eqn{Sdef}{
	S(E,V,N), \qquad
	\pd{S}{E} = \frac{1}{T}, \qquad \pd{S}{V} = \frac{p}{T}, \qquad \pd{S}{N} = -\frac{\mu}{T}.
}

The related  differential form, the Gibbs relation, is: 
\eqn{Gr}{
	\text{d}E = T\text{d}S-p\text{d}V+ \mu \text{d}N.
}
Here \textit{T} is the temperature, \textit{p} is the pressure and \(\mu\) is the chemical potential. The differential form is helpful for the transformation of variables. 
In the following, we assume that the number of particles is constant and entropy is a function of volume and internal energy only. For gases, the usual intensive state functions, equations of state, are the thermal and caloric ones, \(p(\fe,\fv)\) and \(T(\fe,\fv)\). If the entropic intensives, $1/T$ and $p/T$ satisfy that
\eqn{fMr}{
	\frac{\partial}{\partial \fv} \f{1}{T} = 	
	\pd{}{\fe} \f{p}{T},
} 
then there is an entropy function with the partial derivatives given in \re{Sdef}. Up to three variables {\it A2} practically does not require physical conditions, because the temperature is constructed as integrating divisor and the requirement of extensivity, the scaling of thermodynamic properties with size, preserves the potential property for the the two variable function for the third variable, as it was shown in the early analysis of Gyula Farkas \cite{Far895a,Fen52b}. 

The \textit{thermodynamic stability}, {\it A3}, is a further restriction. Then the second derivative of entropy, 
\eqn{fDs}{
	D^2 \fs(\fe,\fv) = \left( 	\begin{matrix} 
		\frac{\partial }{\partial \fe} \frac{1}{T} & 
		\frac{\partial }{\partial \fv} \frac{1}{T}\\
		\frac{\partial }{\partial \fe} \frac{p}{T}& 
		\frac{\partial }{\partial \fv} \frac{p}{T}
	\end{matrix} \right),
}
is negative definite. It is easy to see that it is equivalent to the following inequalities:
\begin{equation}
	\frac{\partial T}{\partial \fe}(\fe,\fv) > 0, \qquad  \frac{\partial p}{\partial \fv}(T,\fv) < 0.
\label{thdcon}\end{equation}
Please note the variables of the second inequality. For an ideal gas, \re{thdcon} is valid in the whole canonical state space, but for a Van der Waals gas, for example, it is not, the region under the so-called spinodal curve in the \((p,\fv)\) diagram violates the second inequality of thermodynamic stability. 

A {\it reservoir} is a thermodynamic body, whose temperature and pressure is constant in any process. Therefore, the  equations of state are constant functions, one can easily calculate the entropy of a reservoir, the \textit{ambient entropy}. Let us denote the constant temperature and pressure of the reservoir by \(T_{0}\) and \(p_{0}\), respectively. Then the respective entropy is 
\begin{equation}
	\fs_{a}(\fe_{a},\fv_{a})= \frac{1}{T_{0}}\fe_{a}+\frac{p_{0}}{T_{0}}\fv_{a}+\fs_{a},
	\label{fskdef}\end{equation} 
where \(\fe_{a}\) and \(\fv_{a}\) are the internal energy, and the volume of the reservoir, respectively, \(\fs_{a}\) is constant. Reservoirs represent idealised, simple environments. 
													
\ssect{hgf_td1}{Termodynamics}
													
Evolution equations of thermodynamics cannot be arbitrary; there are some general restrictions and evident properties of equilibrium that must be fulfilled. For example, the first law  of thermodynamics, i.e. the balance of internal energy, can be related to a differential equation of the following form
\eqn{fkhfv}{
	\dot \fe = q(\fe,\fv,..) + w(\fe,\fv,..). 
}  
Here, the dot denotes the time derivative, \(q\) is the thermal energy transferred per unit time (heating), and \(w\) is the mechanical power, the value of the work done on the homogeneous body per unit time (working). Also, the three dots indicate that the first law is trivial in insulated systems, there the energy is conserved, one expect work and heat exchange only in systems with interactive bodies. The quantities for heat and power are not interpretable unless the thermodynamic body under consideration is in contact with other bodies or at least with its environment. \(q\) and \(w\) are the quantities characterising the interaction. Therefore, they must also depend on the characteristics of the environment or adjacent thermodynamic body or bodies. In the following, we will consider a simple thermodynamic system as shown in Figure \re{fig_egyszr}, a gas in thermal and mechanical interaction with its environment. The dotted spaces in the above formula \re{fkhfv} should be replaced by the characteristic parameters of the environment. 

A particular case is Newton's cooling law, which models the heat exchange of a body and its environment. That example shows some of the expected properties of a complete theory:
\eqn{Newtcool}{
	q(E,V,T_0,p_0) = -\alpha(T(E,V)-T_0).
} 
 $T$ and $T_0$ are the temperatures of the body and the environment, and $\alpha >0$ is the nonnegative heat exchange coefficient. If the mechanical power is zero and the internal energy equals the heat capacity times the temperature, then we can get a simple differential equation for the internal energy. When solved, it gives the time variation of the internal energy, $E(t)$, modelling, e.g. a mug of hot tea in a room temperature environment. However, the combination of heat exchange and mechanical work is not straightforward. 
\begin{figure}\begin{center}
		\includegraphics[width=0.5\textwidth]{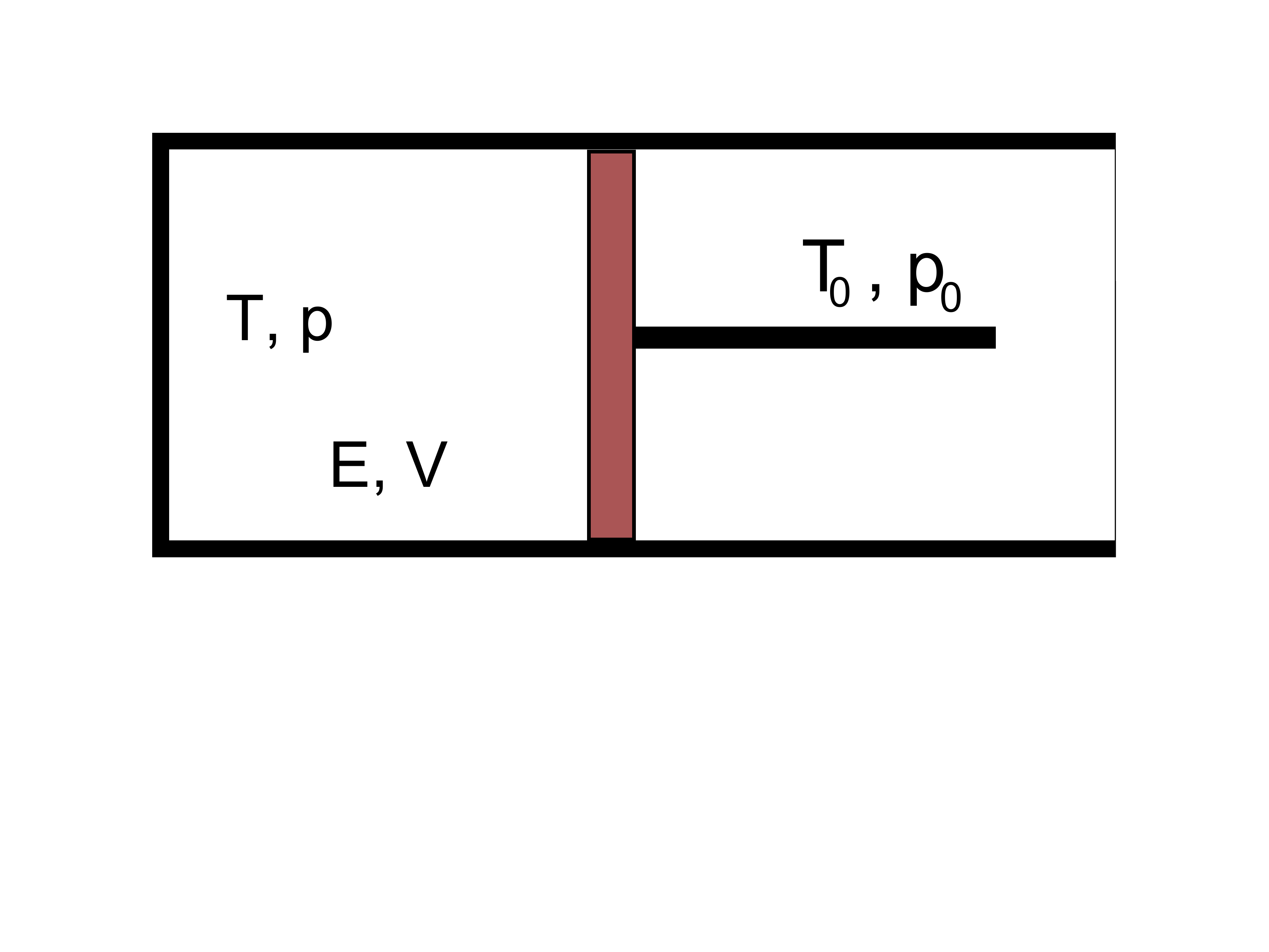}
	\end{center}
	\caption{\label{fig_egyszr} A simple thermodynamic system, a gas connected to a reservoir environment. $T_0$ and $p_0$ are the temperature and the pressure of the reservoir.}
\end{figure}

It is also confusing, that the Gibbs relation, \re{Gr} with $dN = 0$, is associated with the first law. One can find formulas such as
$$
\d E = \delta Q +\delta W = T\d S-p\d V,
$$
where \(\d\) is a differential as before, but \(\delta\) does not just denote some variation, indicating the inconsistent mathematics when combining \re{fkhfv} and \re{Gr}. It is straightforward to reinterpret the above equations with time derivatives:
\begin{equation}
	\dot \fe = q+w \stackrel{?}{=} T\dot \fs -p \dot \fv= \dot \fe. 
\label{G1}\end{equation}

The first equation is the first law, and the last is the time derivative of the internal energy from the Gibbs relation, \re{Gr}. The question is that how could they be valid at the same time? From a mathematical point of view they are unrelated. On the left-hand side, $q$ and $w$ characterise the interaction, depending on both the environment and body characteristics, while on the right-hand side only quantities appear that characterise the thermodynamic body, the gas in the cylinder on Fig. \ref{fig_egyszr}. The only possible consistent explanation requires evolution equations. 

In the following, for the thermodynamic system shown in Fig. \ref{fig_egyszr}, we interpret the first law as the energy balance in the following form:
\eqn{f1ft}{
	\dot \fe = q(\fe,\fv,\fe_{a},\fv_{a}) + w(\fe,\fv,\fe_{a},\fv_{a}). 
}  
However, it is not a complete evolution equation and also the interaction functions are unknown, \cite{TruBha77b}. Therefore for the volume, we introduce
\eqn{fcse}{
	\dot \fv = f(\fe,\fv,\fe_{a},\fv_{a}),
}
where \(f\) characterises the interaction of the body and the environment. Now, one must consider the other expected physical conditions: the requirements of equilibrium, conservation laws and the Second Law. Those restrict the possible forms of the interaction functions, $q,w$ and $f$.  

In a dynamical theory \textit{equilibrium} is time-independence. It is introduced in terms of constitutive functions, usually associated with intensive state variables. In our case, it is expected that when temperatures and pressures are equal, the thermodynamic system does not change; the right-hand side of the above differential equation \re{f1ft}--\re{fcse} is zero. Therefore, it is assumed that the interaction quantities depend on the extensive state variables through the intensive ones and are zero when the intensives are equal:
\eqn{one}{
	q(\fe,\fv,\fe_{a},\fv_{a})&= q\big(T(\fe,\fv), p(\fe,\fv), T_{0}, p_{0}\big), &\text{and}& \quad q\big(T_{0}, p_{0}, T_{0},P_{0}\big)&= 0, \\
	f(\fe,\fv,\fe_{a},\fv_{a})&= f\big(T(\fe,\fv), p(\fe,\fv), T_{0}, p_{0}\big), &\text{and}& \quad f\big(T_{0}, p_{0}, T_{0},P_{0}\big)\! &= 0. 
} 
Note that this seemingly complicated statement is a basic assumption in thermodynamics; it is the part of the zeroth law, \cite{Hon99b,Hsi75b,Cal85b,MulWei05b,Lav10b}.

Mechanical interaction, $w$, is assumed in an ideal form, proportional to volume change, therefore written as
\begin{equation}
	w = - p f.
\end{equation}

Then, the equilibrium solution of the evolution equation, \re{f1ft}--\re{fcse}, is obtained as the solution \((\fe_0,\fv_0)\) of the algebraic equations
\begin{equation}
	T(\fe_0,\fv_0)=T_{0}, \qquad p(\fe_0,\fv_0)= p_{0}.
	\label{egymo0f} \end{equation}  
It is possible that the equilibrium is not unique, for a given constant ambient temperature and pressure \re{egymo0f} may have several solutions. For example a Van der Waals gas equation of state has this property. 

Then we consider that the body-environment system is insulated, therefore
\begin{enumerate}
	\item The volume of the thermodynamical system is constant: \(\fv+\fv_{a} = \fv_{tot}= const.\). From this, it follows that 
	\begin{equation}
		\dot \fv +\dot \fv _{a} = 0.
		\label{vbal0}\end{equation}
	\item The energy of the thermodynamical system is constant: \(\fe+\fe_{a} = \fe_{tot}=const.\). Then
	\begin{equation}
		\dot \fe + \dot \fe _{a} = 0.
		\label{ebal0} \end{equation}
\end{enumerate} 

The entropy of the {\it whole system} is the sum of the entropies of the body and the environment. Its change along the process can be calculated without solving the evolution equation:
\eqn{na1}{
	 \frac{d}{dt}(S+S_{a}) &= 
	\frac{d}{dt}\big(\fs(\fe,\fv) + \fs_{a}(\fe_{a},\fv_{a})\big) = \nnl{na2}
	& = \left(\frac{1}{T}\dot\fe + \frac{p}{T}\dot\fv - 
	\frac{1}{T_{0}}\dot \fe-\frac{p_{0}}{T_{0}}\dot \fv \right)=
	\left(\frac{1}{T}-\frac{1}{T _{0}}\right)(q-pf)+   
	\left(\frac{p}{T}-\frac{p_{0}}{T_{0}}\right)f= \nnl{totentr_deszd}
	&=\frac{1}{T_{0}}\left((T_{0}-T)\frac{q}{T}+(p-p_{0})f\right).    
}
Here we first used the constraints, \re{vbal0}--\re{ebal0}, then the definitions of the entropies of the body and the environment, \re{Sdef} and \re{fskdef}, and finally the evolution equation \re{f1ft}--\re{fcse}. \re{totentr_deszd} is the derivative of the total entropy, $S+S_a$, along a process defined by the system of differential equations \re{f1ft}--\re{fcse}; considering all the processes, $(S+S_{a}\rdot[)]$ will denote the derivative along the evolution equation. According to {\it A4}, the total entropy increases, therefore the interaction function must fulfil the following inequality:
\eqn{2ft0f}{
	(T_{0}-T)\frac{q}{T}+(p-p_{0}) f \geq 0.
} 
This inequality is understood as a constraint on the functions \(q\) and \(f\). Equality can occur only in equilibrium. Its physical meaning is straightforward. For example, in the case of \(f\equiv 0\), it follows from the above inequality that the direction of the heat flow is opposite to the difference between the temperature of the body and the ambient temperature. Therefore \re{2ft0f} is Clausius' formulation of the Second Law,  \cite{Cla879b,Mat05b}. On the other hand, if \(q\equiv 0\), the volume changes in the direction of the pressure difference. That is, if the body's pressure is greater than that of the environment, its volume increases, and if it is smaller, its volume decreases.

There were several conditions -- the concavity of entropy, the dynamic part of the zeroth law for the equilibrium, \re{one}, the growth of entropy, \re{2ft0f} --, and the following statement is a consequence, it is an explanation of their role:

{\bf Stability. }{\it The equilibrium solution, \re{egymo0f}, of the differential equation \re{f1ft}--\re{fcse} is asymptotically stable}.

It is valid, because we can show that \(L(\fe,\fv) = S(\fe,\fv) + \hat S _{a}(\fe,\fv)\) is the Lyapunov function of the equilibrium. Here, \(\hat S _{a}(\fe,\fv) = S _{a}\big((\fe_{tot}-\fe),(\fv_{tot} -\fv)\big)\), i.e. the entropy of the environment expressed by the entropy of the body using the constraints. 

Then, the derivative of \(L\) is zero in equilibrium, and its second derivative is equal to the second derivative of the entropy of the body, which is concave by the requirement of thermodynamic stability. That is, \(L\) has a strict maximum in thermodynamic equilibrium.

Moreover, $\rdot[L]   $ has a strict minimum according to the inequality \re{2ft0f}, hereby the asymptotic stability, according to 

Therefore, the usual assumptions of thermodynamics, in particular  
\begin{itemize}
	\item the existence of entropy as a thermodynamic potential,
	\item thermodynamic stability, i.e. the concavity of entropy,
	\item the interpretation of the entropy of the environment by the conservation constraints, 
	\item the equilibrium conditions,
	\item classical working and
	\item nonnegative entropy production rate, as a condition for interactions
\end{itemize}
together result in the conditions of Lyapunov's theorem of asymptotic stability, the stability and attractivity of the thermodynamic equilibrium (see \cite{Lya892t} or e.g. in \cite{RouEta77b}, theorem 6.2.). 

There are some interesting consequences:

\begin{enumerate}
	\item The Lyapunov function multiplied by the temperature of the environment is the {\it exergy}, the maximum available work of the system, $T_0 \hat S(\fe,\fv) =T_0 S(\fe,\fv) - \fe - p_0 \fv$. Therefore exergy is not a fundamental concept.
	
	\item The derivative of the body entropy along the evolution equation, the entropy production of the thermodynamic body, is the heating divided by the temperature:
	\eqn{rderentr0f}{
		\rdot[\fs] = D \fs\cdot(q-pf,f) =
		\frac{1}{T}(q-pdf) + \frac{p}{T}f = 
		\frac{q}{T}. 
	}   
	Now, the following equation is well-defined:
	\eqn{rds0f}{
		\dot \fe = q+w = T \rdot[\fs] -p \dot \fv.
	}
	Therefore, the relation of the first law and the Gibbs relation, \re{G1}, is clarified.
	
	Usually \(dS = \delta Q/T\) defines the so-called {\em reversible} state changes in thermodynamic textbooks. Now, we are dealing with well-defined processes, and $q/T$ is the derivative of the entropy of the body along the differential equation \re{f1ft}--\re{fcse}. 
	\item The above evolution equation leads to the interpretation of the concept of {\it quasi-static process}. Namely, if the temperature and pressure of the environment are equal to the temperature and pressure of the body at any moment, the body is in equilibrium, i.e. the energy and volume changes cease, and the process stops. In this sense, the process passes through a series of thermodynamic equilibria; it is quasi-static. The nomination quasi-static is misleading and paradoxical, perhaps it is better to call that property {\it controllable}. The instant equilibration indicates the lack of memory and inertia in the system. 
\end{enumerate}

The simple thermodynamic system and evolution equation can be extended to several thermodynamic state variables, bodies and substances: the previous considerations and conditions regarding Lyapunov stability provide a well-defined mathematical framework: ordinary thermodynamics is nonequilibrium thermodynamics of homogeneous bodies. 

From our point of view, the clear physical meaning is the most remarkable: thermodynamics is apparently a theory of stability in a dynamic sense. It is important to emphasise that stability is not only a part of thermodynamics, as it is treated in classical irreversible thermodynamics, \cite{GlaPri71b,KonPri14b}, the theory itself organises the experience establishing a bunch of concepts to construct dynamics with asymptotically stable equilibrium. Asymptotic stability is the central notion of thermodynamics.  

Haddad and coworkers also recognised the role of Lyapunov stability and the possibility of rigorous treatment, \cite{HadEta09b,Had19b}. They indicate, that thermodynamic concepts can be fruitful in optimisation and control and also the mathematics of control theory clarify and refine themodynamical considerations. 

Now we investigate the concept of absolute temperature in the light of the stability framework.

\section{Absolute and universal}

Temperature is a useful concept only if it is universal. Otherwise the temperatures of bodies with different materials could not be compared.  That was the motivation of Lord Kelvin, who was William Thomson then, to introduce absolute temperature, \cite{Tho849a}. The adjective absolute refers to the property of the temperature that it must be independent from material\footnote{It is sometimes referred to as absolute temperature scale. It is not about absolute zero temperature; several recent textbooks and Wikipedia are misleading.}.  

\begin{figure}\begin{center}
		\includegraphics[width=0.5\textwidth]{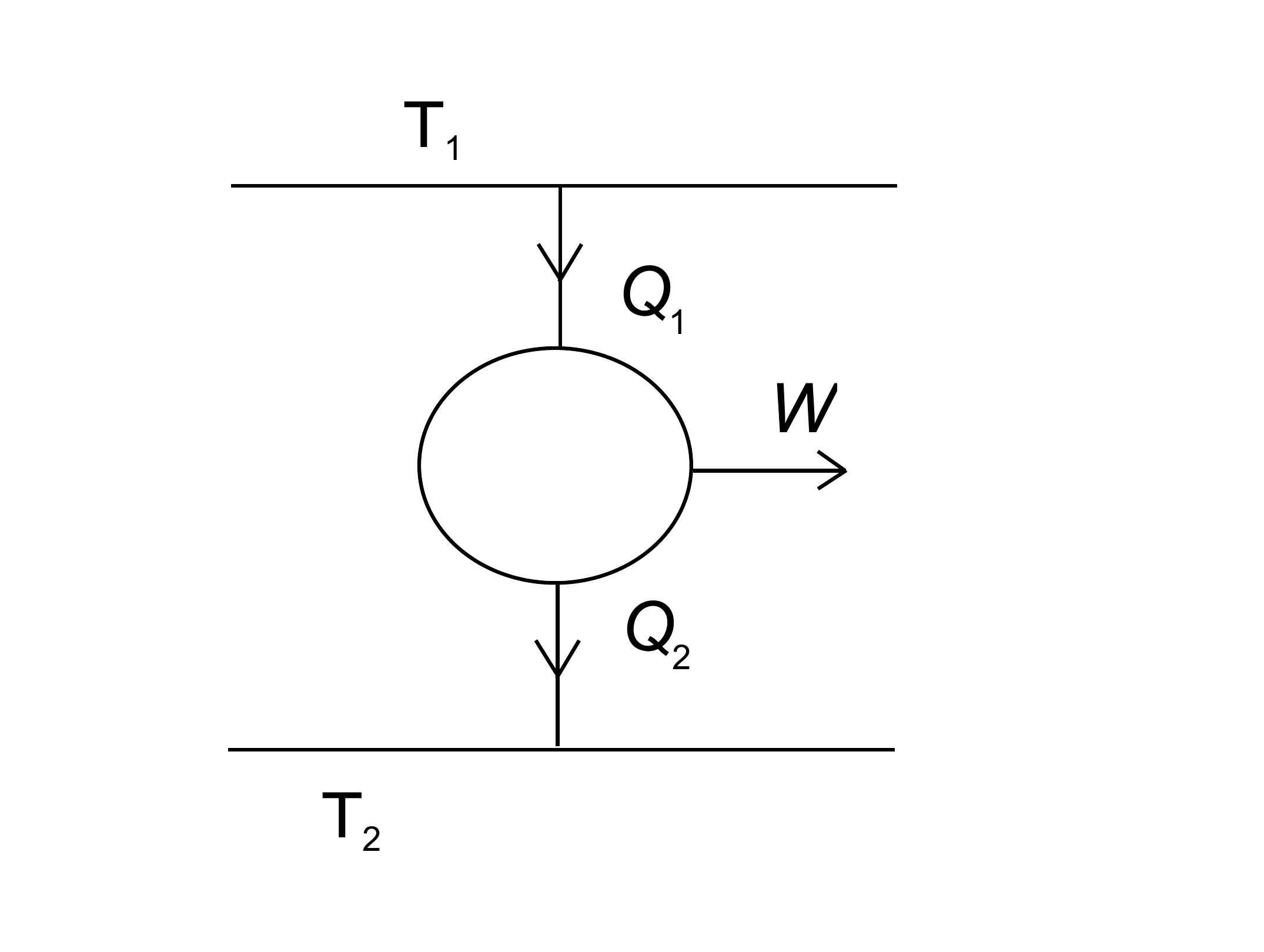}
		\includegraphics[width=0.49\textwidth]{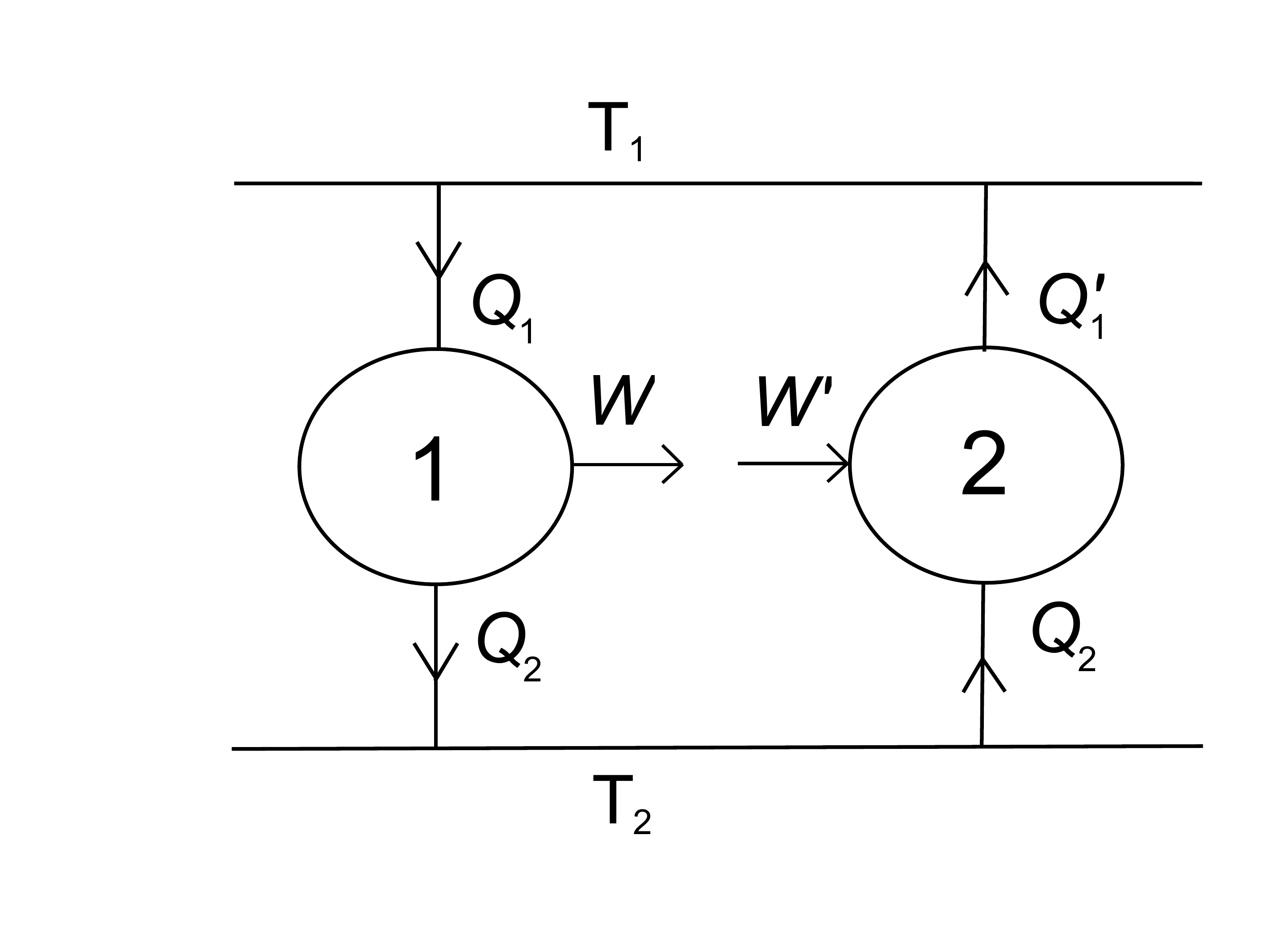}
	\end{center}
	\caption{\label{thd_KP} a) The schematic representation of the heat engine, a thermodynamic system composed of a body, to thermal reservoirs with temperatures $T_1$ and $T_2$ and some other non specified parts absorbing and emitting mechanical work. $Q_1$, $Q_2$ and $W$ are the heat and work emitted and absorbed in a cycle. b) Two engines combined in a forward and reversed mode operation. The work produced by the first engine is used by the second engine to enforce reversed operation. If the second engine is a reversible one and $W>W'$, then the combination of the two becomes a Kelvin-Planck perpetuum mobile.}
\end{figure}

Absoluteness is universality and the related arguments are based on the Kelvin-Planck form of the Second Law and the concept of a reversible engine. In the original reasoning, Second Law is formulated for cyclic processes of a heat engine. In Figure \ref{thd_KP}a, two heat reservoirs with empirical temperatures $T_1 > T_2$ represent constant temperature environments that are connected to the thermodynamic body (the "heat engine" in the classical terminology), where heat exchanges with the reservoirs in one cycle are $Q_1$ and $Q_2$. Also there is mechanical part of the system, therefore mechanical work can be performed and consumed to regulate the operation, e.g. temporarily connect or disconnect the reservoirs, enabling or disabling the heat exchange with the reservoirs. The total mechanical work in a cycle, $W$, is the difference of the adsorbed and emitted heats, $W=Q_1-Q_2$, according to the First Law. The structure of the engine can be arbitrary, and at the end of a cycle period, the state of the engine is restored. The Kelvin-Planck formulation of the Second Law claims that $Q_2$, the emitted heat in a cycle, cannot be zero. No engine could perform work and only absorbs heat in a cycle. There is no perpetuum mobile of the second kind.

It is not assumed that the reservoirs are connected continuously to the engine or would be connected to the engine simultaneously. Then, an analysis of the system reveals that the Kelvin-Planck statement follows from the existence of body entropy as a potential, \cite{TruBha77b,Mat05b}. Also, the maximum efficiency is the Carnot efficiency. 
However, the existence of entropy of the body -- therefore the Kelvin-Planck form -- cannot say anything about the interactions, about heat exchange or about work, including their directions. Neither is assumed that the engine's temperature would be the same as the temperature of the reservoir\footnote{That is the niche, where finite time thermodynamics exist, and the freedom of optimising for maximum power appears, \cite{CurAhl75a}. However, the fundamental importance of the Second Law remains intact. That can be understood also when compared to the original and literature, like the works of Carnot, Kelvin and Maxwell among others, \cite{Lav10b}.}. 

The mechanical power during operation depends on time, there is a function $\pi: t\mapsto \pi(t)$, and $W=\int_{cycle} {\pi(t)dt}$. There are engines, whose operation can be reversed in the sense that there is a power function that regulates the machine adsorbing $Q_2$ heat in a cycle from the reservoir with temperature $T_2$. The required work, $W'$ and the emitted heat to the $T_1$ reservoir $Q_1'$ are different\footnote{A reverse cycle air conditioner, a heat pump, is an example. It is not completely mechanical, requires electricity and there are other differences, but otherwise can be operated forward and backward, in a cyclic mode. However, its original state never will be restored.}. The power function for reversed operation is not necessarily the forward power with negative sign. The timing and duration of reversed operation, opening and closing of thermal and mechanical parts, the insulation and reconnection from and to the reservoirs may be different, too. The engine is called {\it reversible}, if the work required for reversed operation is the same as for forward operation, while adsorbing and emitting the same amount of heat from and to the $T_2$ reservoir in the reversed and forward modes, respectively. 

Reversible operation is somehow more intricate than one may expect. First of all, good thermal engines are not the best heat pumps, because the {\it thermal efficiency} is defined as $\eta = \f{W}{Q_1}$, it is the better the more work is produced from less adsorbed heat. The {\it effectiveness} of a heat pump, assuming a reversed operation on Figure 2/a, is defined as $\eta = \f{Q_2}{W}$, it is the better the more heat is adsorbed by less work. $\eta = \f{1}{1+\epsilon}$, therefore a bad thermal engine is a good heat pump and a good heat pump is a bad thermal engine, if one assumes reversible operation. However, the worst possible heat pump, with $Q_2=0$, would dissipate all work to heat, without adsorbing any heat from the $T_2$ reservoir. That kind of operation cannot be reversible, as follows from the above mentioned Kelvin-Planck form of the Second Law. Also, if I take a heat pump with low effectiveness, I do not necessarily find a thermal engine that would have a larger thermal efficiency than the heat pump in a fictional reversed operation. 

A reversible engine has the largest possible thermal efficiency, it is the best possible thermal engine. The classical reasoning is seemingly simple (see e.g. in \cite{Max902b}): if any machine has larger efficiency than the reversible one, it would be easy to construct a perpetuum mobile of the second kind. If the thermal efficiency of the forward engine, $\eta = \f{W}{Q_1}$ is larger than the efficiency of the reversible one, $\eta' = \f{W'}{Q_1'}$, then the first combined with the reversible one operated in a reverse mode becomes a perpetuum mobile, because at he end of the cycle the exchanged heat with $T_2$ is zero, and  $1-\eta = \f{Q_2}{Q_1} < 1-\eta' = \f{Q_2}{Q_1'}$, therefore $Q_1'< Q_1$ and $W'<W$. The combined engine adsorbs $Q_1- Q_1'$ heat from the $T_1$ reservoir while performing $W-W'$ work. Therefore the reversible engine has the largest possible efficiency not violating the Second Law. However, the simplicity is misleading: a reversible engine in a reverse operation is not the worst possible heat pump, one cannot argue with reversible heat pumps instead of reversible thermal engines. It is because mechanical work can be converted to heat without problems, and the extra work reduces the efficiency of the machine. On the other hand, if that dissipated mechanical work is directed toward the $T_1$ reservoir, contributing $Q_1$, then it can be useful as direct heating. One should be careful not mixing technical requirements with theoretical concepts.

Thermal efficiency can depend only on the temperatures of the reservoirs. If the difference in construction and materials leads to a difference in efficiency, then one can combine the two like in the previous case; the one with a smaller efficiency is operating backwards the same perpetuum mobile could be constructed. Therefore the efficiency is universal and can be used to define the absolute temperature, \cite{Tho849a,Max902b,Kar07b1,Lav10b}. 

The Second Law and the reversible operation are the most apparent conditions in the the argumentation above. The reversible operation of a reversible engine does not require step-by-step reversibility of the corresponding processes, the forward and reverse power functions can be different. The possibility of cyclic operation is trivial from a technical point of view. It must be ideal in the sense that the engine must recover its original state at the end of the cycle. However, a reversible engine can operate in a reverse mode while the heat and work exchanges are the same as in the case of forward operation but with the opposite sign. 

In the simplest case, if the engine is a single thermodynamic body with the evolution equations of the previous section, the recoverability of the original state is provided if the contact with the reservoir and the original external pressure is restored. Then the original equilibrium is recovered because the processes are controllable in the sense of the previous section. For simple equations of state, e.g. for an ideal gas, complete control of the engine is provided by the regulation of the external pressure, $p_a(t)$, a Carnot process could run either forwards or backwards, and the conditions of well defined universal temperature are fulfilled. 

In summary, the universality of the temperature is based on the Second Law. 

\section{The role of universality}

Universality in physics has empirical and analytical aspects. The widespread appearance of thermodynamical concepts in sciences and humanities can be considered as indications of empirical universality, \cite{Had17a}. One may look for common background, arguing, that general assumptions lead to general consequences. It was demonstrated with the concept of absolute temperature, where the universality was justified by the Kelvin-Planck form of the Second Law. Moreover, the Kelvin-Planck form is based on the existence of entropy as thermostatic potential, it is clear, that it has nothing to do with process directions\footnote{Clausius form of the Second Law is independent of the Kelvin-Planck statement. The requirements regarding the interaction of thermodynamic bodies are independent of the requirements on the material properties of the bodies. It is clear from the stability statements but also can be proved directly, see \cite{TruBha77b,Mat05b}.}. Postulating the existence of a potential is a weak condition, in particular considering the lack of any microscopic background. 


In this paper, it was argued that thermodynamics is a stability theory. It was demonstrated for homogeneous thermodynamic bodies. There the seemingly independent concepts, assumptions and properties of equilibrium thermodynamics can be unified as parts of Lyapunov stability conditions in the framework of a dynamical theory with genuine evolution equations with general assumptions about the material properties or the interactions. Then thermodynamics -- in a general sense -- provides a simple mathematical structure that ensures existence and stability of equilibrium. The evolution equation was constructed according to thermodynamics, and thermodynamic conditions were well suited into the stability framework. Remarkably, the differential equation for the volume change was constructed using only stability arguments. If the conditions, like the fundamental balances, are adequately considered in the stability calculations, i.e. in the specification of the state space and in the exploitation of the entropy balance, then the theory, including the evolution equations, is universal. It is as universal as general the assumptions are. 

The existence and stability of equilibrium is the most general assumption that one may expect for any natural evolution. Without stability, objective, repeated experiments are impossible. The unstable state of matter transforms into a stable one. Therefore a constructive approach where the only requirement is the stability of equilibrium leads to universal evolution. That is also valid for evolution in spacetime. Therefore, it is not surprising that evolution equations derived with the help of a new constructive methodology of nonequilibrium thermodynamics are robust models in heat conduction or continuum mechanics \cite{KovVan15a,SudEta21a}, that thermodynamic concepts emerge concerning the most fundamental theories of physics, \cite{Ver11a}. For example the field equation of Newtonian gravity emerges without any further ado, \cite{VanAbe22a}. The Second Law and a pure thermodynamic based stability framework looks like fundamental, beyond expectations, \cite{VanKov20a,Van22m}. 

It is remarkable, that any structural interpretation, e.g. particle-based argument of statistical physics, may destroy universality but not necessarily do that. If the statistical concept is compatible with the stability structure, e.g. defines the thermodynamic state variables, then thermodynamics becomes emergent. Also, at the same time, an inconsistent microscopic identification of the macrovariables necessarily reduces the universality. It is remarkable that evolution equations of probability distribution functions, like the Boltzmann equation, can be interpreted in a thermodynamic framework, too, \cite{PavEta18b}.

The Second Law is expected to play a role in various disciplines of science, but its application in the case of economics, ecology, biology or anywhere else is somehow mysterious. Random behaviour of individual objects alone does not justify the emergence of thermodynamic quantities like energy or entropy, nor the tendency of its uniform distribution. However, thermodynamics provides a straightforward and natural framework when some form of equilibrium exists, and stability is expected or required. It is not a wonder that growth models of economics, structural stability of ecology, and dynamical aspects of evolutionary game theory are similar to thermodynamic evolution, \cite{SzaBor16a,BirNed20a}. There are still several steps towards a universal theory of stable evolution.

\section{Acknowledgement}   
The work was supported by the grants National Research, Development and Innovation Office –  FK134277. The research reported in this paper is part of project no. BME-NVA-02, implemented with the support provided by the Ministry of Innovation and Technology of Hungary from the National Research, Development and Innovation Fund, financed under the TKP2021 funding scheme.

The author thank the referees for their valuable and detailed comments and remarks. 

\bibliographystyle{unsrt}

\end{document}